\documentclass[a4paper,11pt]{article}
\usepackage{pos}

\title{Calorimetry for the ePIC Experiment}

\author*[a]{Henry T. Klest}
\affiliation[]{On behalf of the ePIC collaboration\newline}
\affiliation[a]{Argonne National Laboratory,\\
  9700 S. Cass Ave., Lemont, IL, USA}

\emailAdd{hklest@anl.gov}

\abstract{The Electron-Ion Collider (EIC) will deliver collisions of electrons with protons and nuclei at a wide variety of energies and at luminosities up to 1000 times higher than HERA. Precise measurement of both the scattered electron and the hadronic final state is crucial for the physics of the EIC, necessitating unique designs for the electromagnetic and hadronic calorimeters in the backward (-4 < $\eta$ < -1.4), central (-1.4 < $\eta$ < 1.4), and forward (1.4 < $\eta$ < 4) regions. To ensure maximal containment of energy and acceptance for the required physics processes, the Electron-Proton/Ion Collider (ePIC) detector employs calorimetry over almost the entire polar angle. This proceedings will provide an overview of the current calorimeter designs being employed in ePIC.}

\FullConference{31st International Workshop on Deep Inelastic Scattering (DIS2024)\\
 8–12 April 2024\\
Grenoble, France\\}

\begin{document}
\maketitle

\section{Calorimetry at the EIC}
\begin{figure}[htbp]
\begin{center}
\includegraphics[width=15cm]{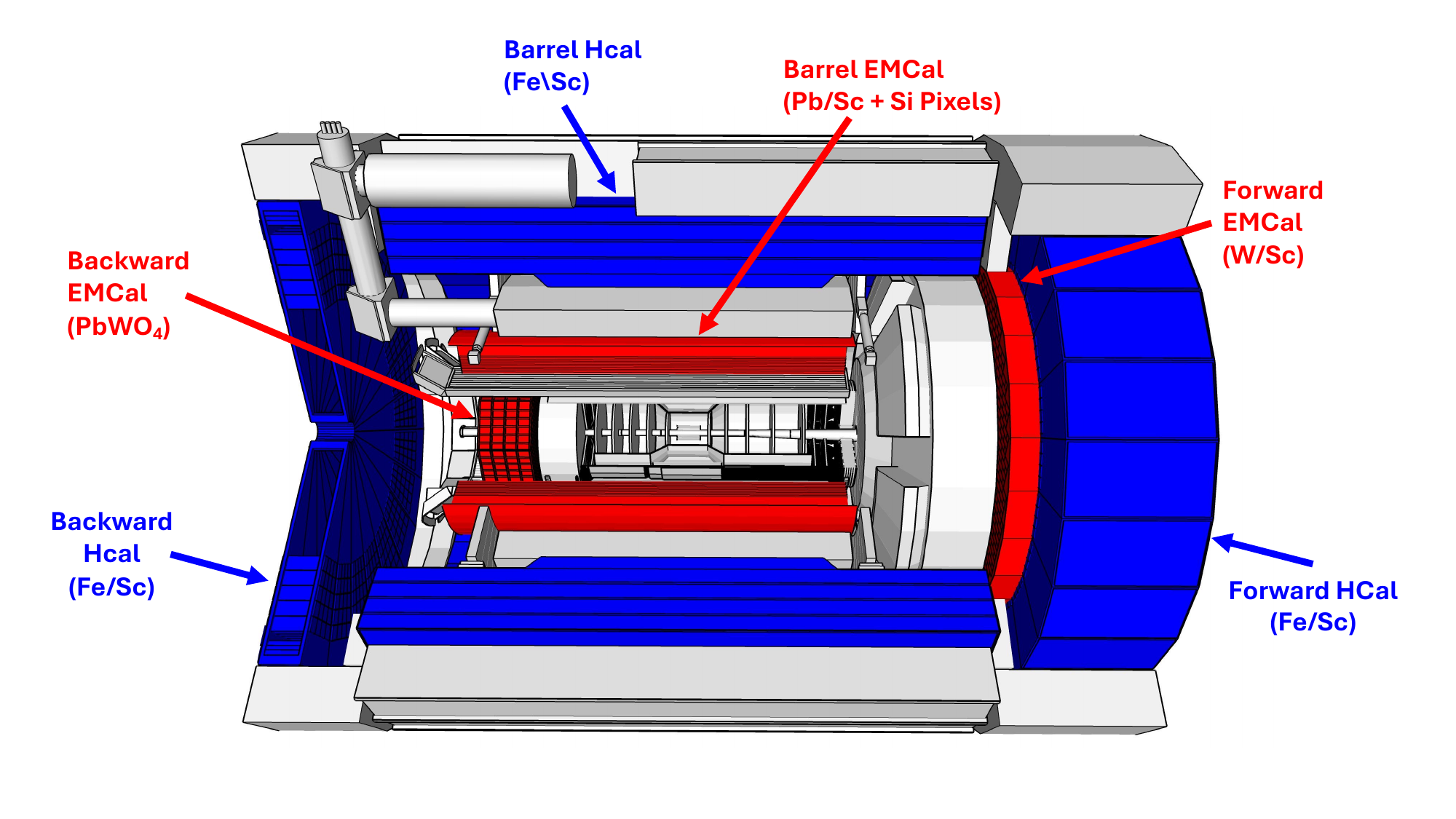}
\caption{Overview of the ePIC central detector. Electromagnetic calorimeters are colored red, and hadronic calorimeters are colored blue.}
 \label{fig:CaloOverview}
\end{center}
\end{figure}

The ePIC experiment is a general-purpose, hermetic collider detector with the goal of carrying out the broad EIC physics program~\cite{AbdulKhalek:2021gbh}. The EIC will collide electrons with protons and nuclei at center of mass energies ranging from $\sqrt{s}\approx30$ GeV to $140$ GeV, at luminosities up to $10^{34}\text{ cm}^{-2}$ per second~\cite{EICParameters}. The physics of the EIC imposes stringent requirements on tracking, particle identification, and calorimetry, with each region of the detector subject to unique challenges. A few of the challenges facing the calorimeter systems (shown in Fig.~\ref{fig:CaloOverview}) are:
\begin{itemize}
\setlength\itemsep{-0.4em} 
    \item Identifying and precisely measuring the scattered electron.
    \item Measuring single particles with momenta from tens of MeV to tens of GeV.
    \item Containing jets with energies over 100 GeV and providing information to particle-flow reconstruction algorithms.
    \item Separating single photons from the two photons arising in decays of neutral pions.
\end{itemize}
In addition to these physics requirements, the calorimeters must be able to handle streaming readout at up to 500 kHz of event rate and radiation loads of $5\cdot10^9$ (in the backward and barrel regions) to $2\cdot10^{11}$ (in the forward region) 1 MeV neutron equivalent dose per cm$^{2}$ per year.



\section{ePIC Electromagnetic Calorimeters}
\subsection{Backward Electromagnetic Calorimeter}
\begin{figure}[htbp]
\begin{center}
\includegraphics[width=14cm, trim=0cm 0cm 0cm 2.8cm, clip]{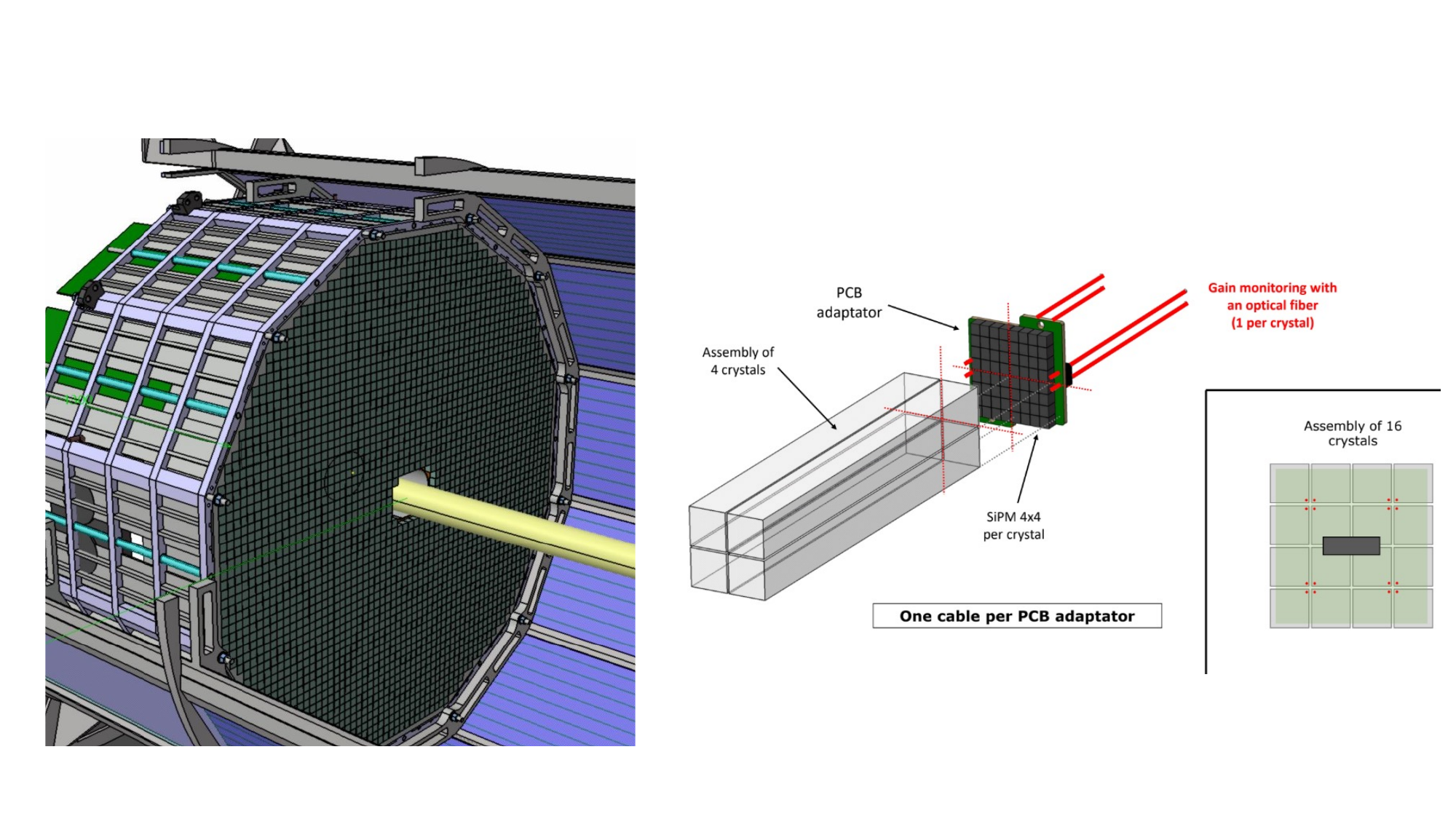}
\caption{Left: CAD drawing of the backward ECal integrated in ePIC. Right: Readout and LED monitoring scheme.}
 \label{fig:BackwardECal}
\end{center}
\end{figure}

The design of the backward electromagnetic calorimeter is driven in large part by the requirement of excellent energy resolution for measuring the scattered electron kinematics and separating pion showers from electron showers via E/p. To reduce the large photoproduction background for DIS observables, the rate of pions being misidentified as electrons should be on the order of 1-in-10000 or better for the combined system of tracking, PID, and calorimetry. The energy resolution required for the calorimeter is on the order of $\frac{\sigma_E}{E} \approx \frac{2\%}{\sqrt{E}}\oplus1-3\%$. Furthermore, the design should be radiation hard, have a small Moli\`ere radius, and enable detection of photons above 50 MeV. The only realistic option for meeting all of these requirements is a homogeneous calorimeter based on scintillating lead tungstate crystals. The backward ECal design, shown schematically in Fig.~\ref{fig:BackwardECal}, is similar to that of the Neutral Particle Spectrometer currently taking data in Hall C at Jefferson Lab~\cite{Horn:2015yma,Horn:2019beh}. The crystals are rectangular with dimensions 2x2x20 cm$^3$, resulting in around 22 $X_0$ for particles at normal incidence. The PbWO$_4$ scintillation light from an individual crystal is measured by a 4x4 array of Hamamatsu S14160-3010PS SiPMs. To reduce the amount of dead material between the crystals, they are supported by two thin (0.5 mm) frames of carbon fiber located at the front and back of each crystal.

\subsection{Barrel Electromagnetic Calorimeter}
Similar to the backward direction, the barrel region of ePIC should deliver an electron-pion separation power of 1-in-10000. To achieve this level of separation, the ePIC barrel electromagnetic calorimeter, also known as the Barrel Imaging Calorimeter or BIC, incorporates a lead/scintillating fiber design with HV-MAPS AstroPix~\cite{Brewer:2021mbe,Steinhebel:2022ips} silicon pixel detectors designed for the AMEGO-X~\cite{Kierans:2020otl} gamma-ray astronomy missions. The barrel calorimeter is around 4.5 meters long including services, is divided azimuthally into 48 sectors, and covers pseudorapidities between -1.7 and 1.3. The Pb/SciFi section emulates the design implemented successfully in GlueX~\cite{Beattie:2018xsk} and KLOE~\cite{Adinolfi:2002zx}. The Pb/SciFi bulk section of the calorimeter utilizes 435 cm-long scintillating fibers readout on both sides by 1.2 cm x 1.2 cm Hamamatsu S14161 SiPM arrays of 50 micron pixel size, of which there are 60 per sector per side. The AstroPix sensors consist of 500 $\mu$m x 500 $\mu$m square pixels capable of measuring dE/dx via time-over-threshold. The low power consumption of AstroPix, on the order of a few mW/cm$^2$, enables them to be inserted between the layers of Pb/SciFi without substantial cooling infrastructure. The high spatial granularity of the AstroPix sensors, combined with the excellent energy resolution of the SciFi portion, enables this detector to meet the strict electron-pion and $\pi^0/\gamma$ separation requirements. The energy resolution is expected to be $\frac{\sigma_E}{E} \approx \frac{5\%}{\sqrt{E}}\oplus1\%$.
\begin{figure}[htbp]
\begin{center}
\includegraphics[width=13cm, trim=0cm 3cm 0cm 2.8cm, clip]{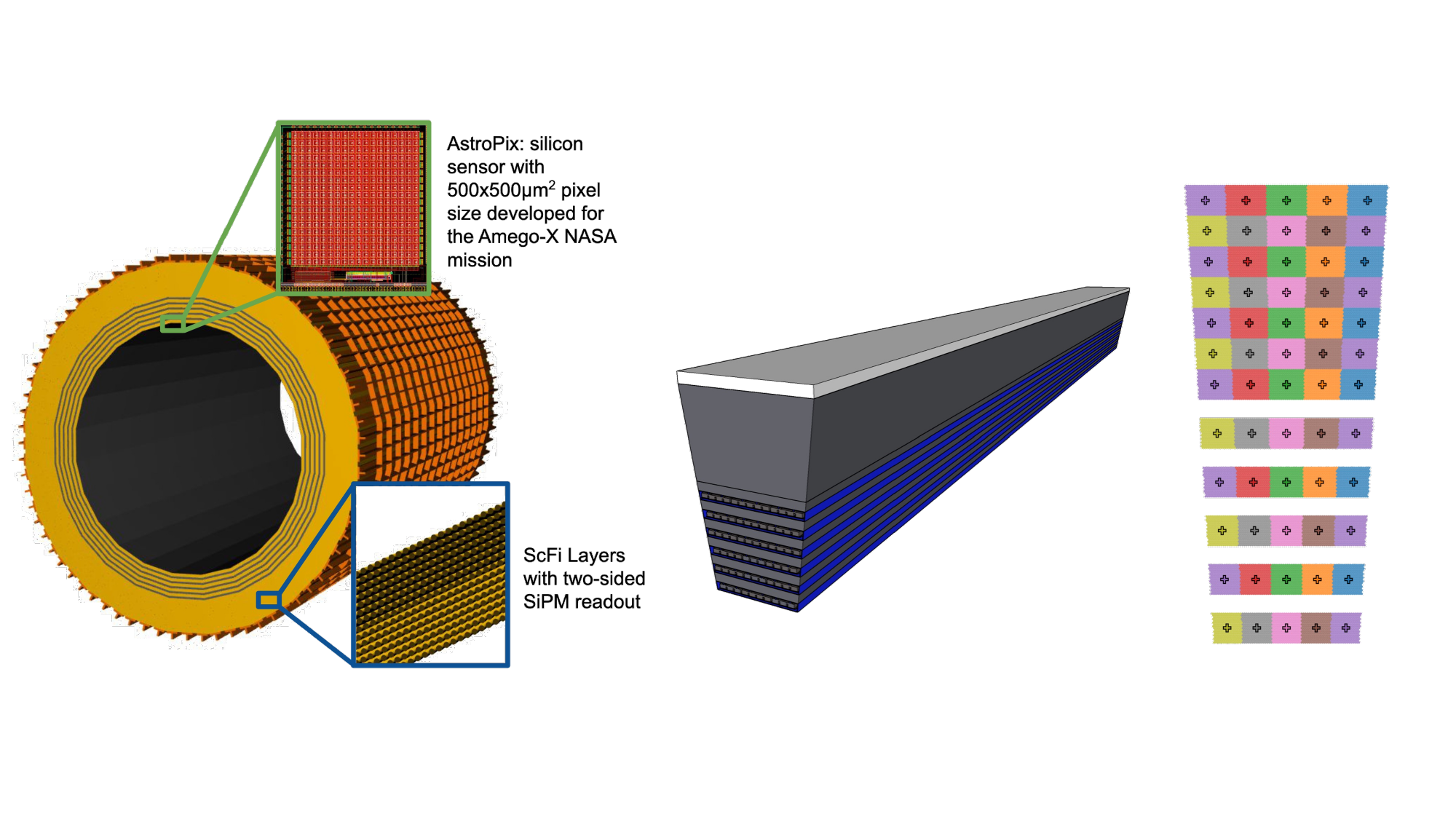}
\caption{Left: Schematic of the full barrel ECal. Middle: Model of a single sector of the barrel ECal, gray denotes the Pb/SciFi bulk, the slots which contain the AstroPix are highlighted in blue. Right: Readout scheme of the Pb/SciFi section. Each trapezoidal tile corresponds to a single readout channel.}
\label{fig:BarrelECal}
\end{center}
\end{figure}

\subsection{Forward Electromagnetic Calorimeter}
The ePIC forward electromagnetic calorimeter builds on the Tungsten/SciFi SpaCal design studied throughout the EIC R\&D effort~\cite{Tsai:2015bna,Tsai:2012cpa} and applied in the sPHENIX experiment~\cite{sPHENIX:2017lqb,Aidala:2020toz}. The energy resolution requirements are modest, but the detector should have the granularity and density necessary to reduce the number of high-energy $\pi^0$ decay photon pairs being misreconstructed as single photons. The scintillating fibers run in the z-direction and are embedded in a mixture of tungsten powder and epoxy, which provides an overall density of around 10 g/cm$^2$. A block of W/SciFi is 5 x 5 x 17 cm and is subdivided into four towers. Each tower is instrumented with four 6x6 mm Hamamatsu S14160 series SiPMs. The light is guided from the face of the block to the SiPM by a 2 cm light guide. This design allows for separation of $\pi^0$ and photon clusters up to around 40 GeV. The expected energy resolution of the forward EMCal is $\frac{\sigma_E}{E} \approx \frac{10\%}{\sqrt{E}}\oplus1-3\%$.

\section{ePIC Hadronic Calorimeters}
\subsection{Backward Hadronic Calorimeter}
The hadronic final state at low-$x$ is typically scattered in the backward direction, necessitating the ability to distinguish the scattered electron from energy deposits created by the final state hadrons. This ambiguity limited the precision of HERA measurements at low-$x$. Since the energies of hadrons in the backward direction are not very high, the backward hadronic calorimeter serves primarily as a tail-catcher for hadrons and a muon identification system for decays of vector mesons. The backward hadronic calorimeter subtends the region $-4.1 < \eta < -1.2$. The design consists of ten layers of 4 cm thick non-magnetic steel and 4 mm thick plastic scintillator, producing a total depth of around 2.4 $\lambda_0$. 
\subsection{Barrel Hadronic Calorimeter}
\begin{figure}[htbp]
\begin{center}
\includegraphics[width=12cm, trim=2cm 4cm 2cm 2cm, clip]{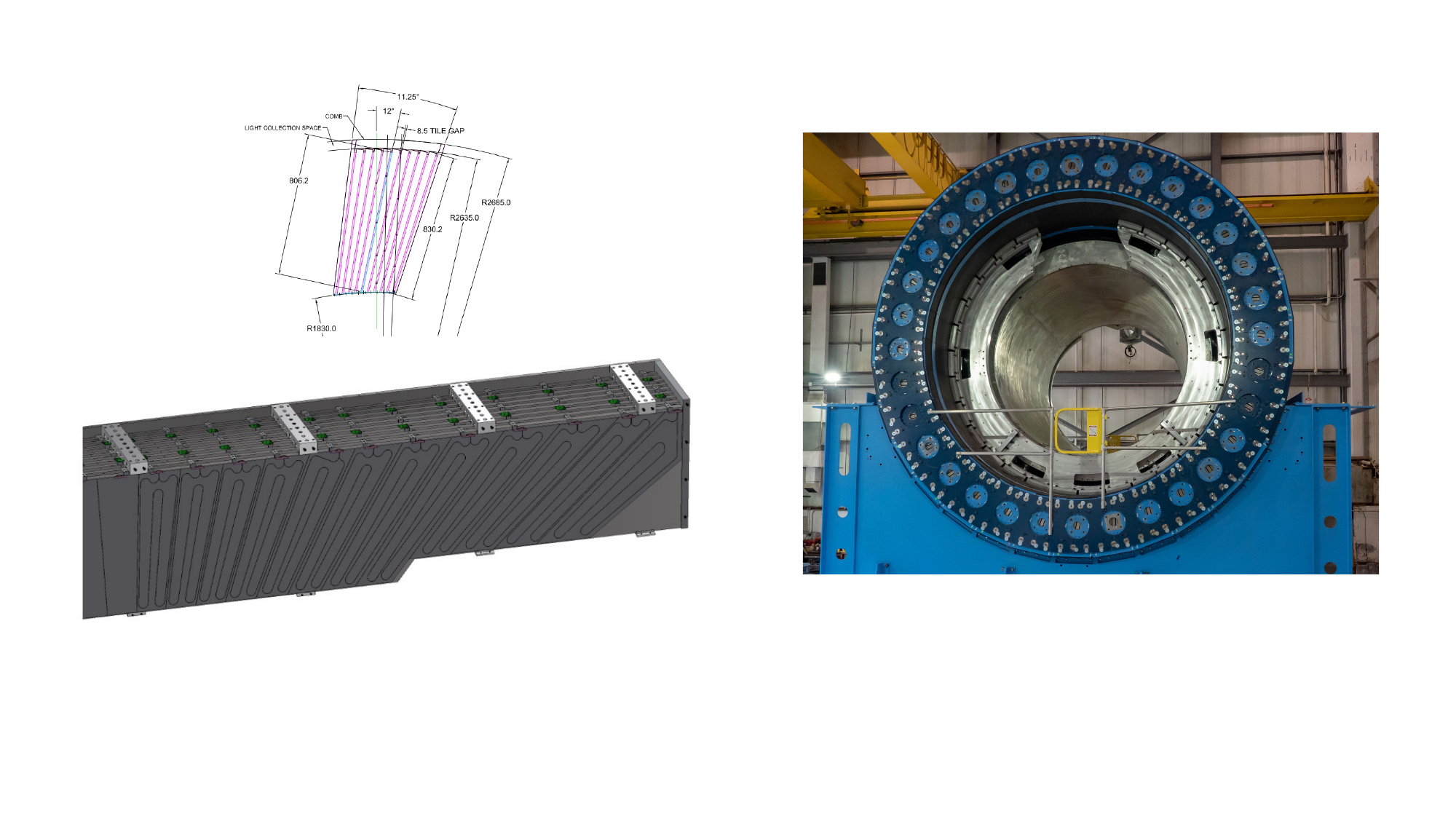}
\caption{Top Left: Side view of a barrel HCal sector, demonstrating the 12 degree tilt. Bottom Left: Rendering of a BHCal sector; the path of the wavelength shifting fiber through the scintillating tile can be seen. Right: Outer HCal (blue) supporting the sPHENIX magnet (silver).}
 \label{fig:BarrelHCal}
\end{center}
\end{figure}

The ePIC barrel hadronic calorimeter is a refurbished version of the sPHENIX outer HCal, described in Refs.~\cite{sPHENIX:2017lqb,Adkins:2022jfp}. The absorber consists of long magnetic steel plates tilted by 12 degrees in azimuth, as shown in the top left portion of Fig.~\ref{fig:BarrelHCal}. Between the steel plates are 7 mm thick scintillating tiles, in which are embedded wavelength shifting fibers to collect the scintillation light and transport it to an SiPM on the outer radius of the detector. The 12 degree tilt ensures that a particle travelling radially will hit at least four scintillating tiles. The expected energy resolution for single hadrons is around $\frac{\sigma_E}{E} \approx \frac{75\%}{\sqrt{E}}\oplus15\%$.

\subsection{Forward Hadronic Calorimeter}
\begin{figure}[h]
\begin{center}
\includegraphics[width=12cm, trim=0cm 0cm 0cm 0cm, clip]{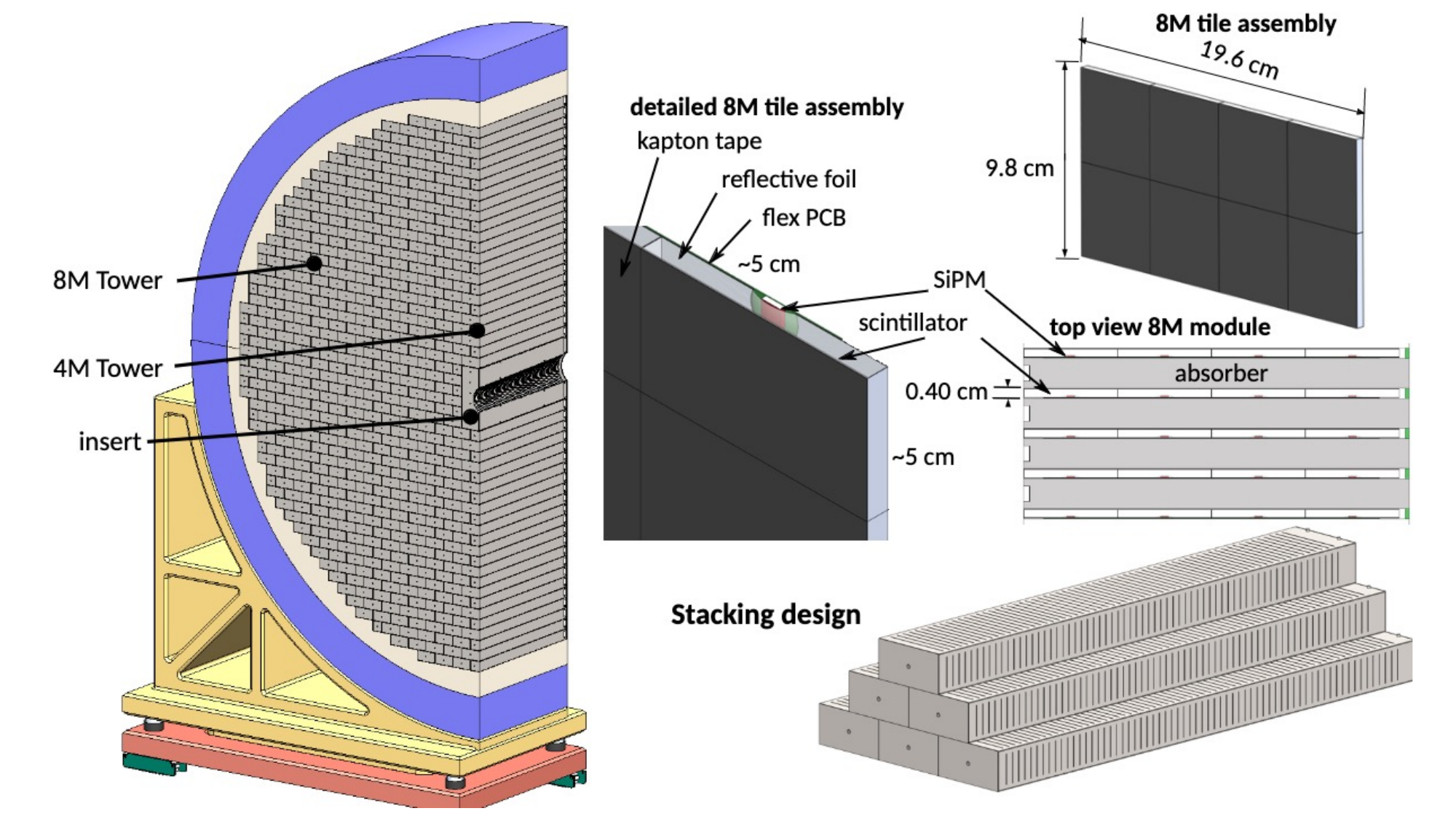}
\caption{Design of the LFHCal. Left: Drawing of one-half of the LFHCal on its support structure. Right: Layout of an LFHCal module.}
 \label{fig:FwdHCal}
\end{center}
\end{figure}
The ePIC forward hadronic calorimeter, known as the Longitudinally-segmented Forward HCal (LFHCal), is designed to contain and precisely measure the high-energy hadrons produced in the forward region. The design leverages the SiPM-on-tile technology pioneered by the CALICE AHCal~\cite{CALICE:2010fpb} to provide fine spatial granularity, ideal for particle-flow algorithms. Each tower consists of steel absorber interleaved with 65 layers of eight square scintillator+SiPM tiles (See Fig.~\ref{fig:FwdHCal}). The SiPM signals are ganged longitudinally into 7 readout channels, resulting in 56 readout channels per tower. The full detector contains 565,760 SiPMs readout via 60,928 channels of the HGCROC ASIC, developed for the CMS High Granularity Calorimeter~\cite{Gonzalez-Martinez:2023wee}. The LFHCal achieves an excellent energy resolution of around $\frac{\sigma_E}{E} \approx \frac{44\%}{\sqrt{E}}\oplus6\%$. The innermost radius of the forward HCal, where the radiation load and particle energies are highest, consists of an "insert" section instrumented with hexagonal scintillating tiles. Detailed information on the insert section can be found in Refs.~\cite{Arratia:2022quz,Arratia:2023xhz}.

\section{Summary}
\begin{figure}[h!]
\begin{center}
\includegraphics[width=13cm]{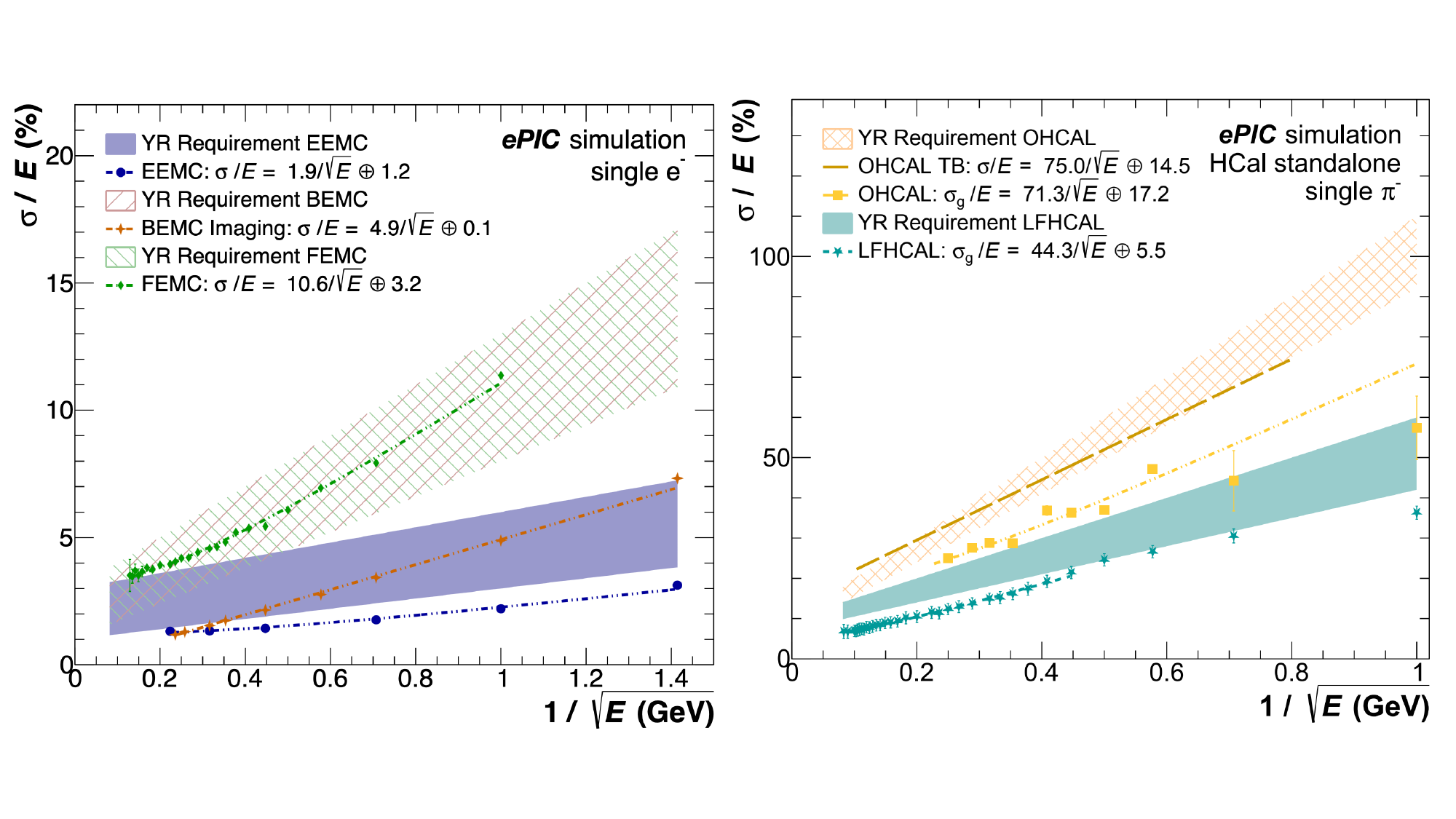}
\caption{Left: Electromagnetic calorimeter resolution projections compared to the Yellow Report requirements. EEMC refers to the backward EMCal, BEMC refers to the barrel EMCal, and FEMC refers to the forward EMCal. Right: Hadronic calorimeter resolutions compared to the Yellow Report requirements. OHCal refers to the barrel HCal, and LFHCal refers to the forward HCal.}
 \label{fig:EnergyRes}
\end{center}
\end{figure}
In summary, the ePIC calorimeter systems meet or exceed the challenging energy resolution requirements laid forth in the EIC Yellow Report, as can be seen in Fig.~\ref{fig:EnergyRes}. The ePIC collaboration is presently in the process of writing a technical design report that will provide detailed designs and projected performances of the calorimeters and other detector subsystems.
\bibliographystyle{plain}
\bibliography{bib}
\end{document}